\documentclass[aps,pra,twocolumn,groupedaddress,bibnotes,
 amsmath,amssymb]{revtex4-1}

\usepackage{graphicx}
\usepackage{dcolumn}
\usepackage{bm}
\usepackage{hyperref}

\usepackage{changes}

\newcommand{\kaex}{\kappa^{(ex)}_{a}}
\newcommand{\kbex}{\kappa^{(ex)}_{b}}

\newcommand{\nelth}{\overline{n}_\mathrm{el,th}}

\begin{document}

\title{On-chip microwave-to-optical quantum coherent converter based on a superconducting resonator coupled to an electro-optic microresonator}

\author{C.~Javerzac-Galy}
\author{K.~Plekhanov}
\author{N.~R.~Bernier}
\author{L.~D.~Toth}
\author{A.~K.~Feofanov}
\email[]{alexey.feofanov@epfl.ch}
\author{T.~J.~Kippenberg}
\email[]{tobias.kippenberg@epfl.ch}

\affiliation{\'{E}cole Polytechnique F\'{e}d\'{e}rale de Lausanne (EPFL), CH-1015 Lausanne, Switzerland}

\date{\today}

\begin{abstract}
 We propose a device architecture capable of direct quantum electro-optical conversion of microwave to optical photons. 
 The hybrid system consists of a planar superconducting microwave circuit coupled to an integrated whispering-gallery-mode 
 microresonator made from an electro-optical material. 
 We show that by exploiting the large vacuum electric field of the planar microwave resonator, electro-optical (vacuum) coupling rates $g_0$ as large as $\sim 2\pi \, \mathcal{O}(10-100)$~kHz are achievable with currently available technology -- a more than three order of magnitude improvement over prior designs and realizations. 
Operating at millikelvin temperatures, such a converter would enable high-efficiency conversion of microwave to optical photons. 
 We analyze the added noise, and show that 
 maximum quantum coherent conversion efficiency
 is achieved for a multi-photon cooperativity of unity
 which can be reached with optical power as low as $ \mathcal{O}(1)\,\mathrm{mW} $.
\end{abstract}


\maketitle

The interconversion of microwave and optical signals 
is of practical relevance in a broad range of electronic applications, 
from optical and wireless communications to timing. 
The spectacular advances of the past decade 
in manipulating the quantum states of the microwave field \cite{Hofheinz2009, Devoret2013}
has increased interest in techniques to convert them to optical fields,
 since the latter can be propagated via optical fiber
 at room temperature 
  while preserving their quantum state.
In the long term, converting quantum states between microwave and optical photons 
may enable long distance quantum communication 
\cite{Barzanjeh2012, Kimble2008},
and in the near term,  it provides 
a path towards realizing single photon detectors of the microwave field that may find use in quantum science and metrology, radio astronomy and technology alike. 
For these reasons, hybrid systems 
for such microwave to optical interfaces 
have recently attracted significant experimental efforts. 
Several approaches have been investigated \cite{Tian2015, Schoelkopf2008}: 
optomechanical and electromechanical devices 
\cite{Bochmann2013, Bagci2014, Andrews2014, Pirkkalainen2013} 
as well as cold atoms \cite{Hafezi2012} 
and spin ensembles \cite{Kubo2010, Longdell2014}. 
Indeed, a bi-directional and efficient link has been established recently 
using a mechanical oscillator coupled to both optical and microwave modes. 
Alternatively, it has been proposed that the parametric coupling of 
an LC circuit to an optical cavity 
via an electro-optical crystal would
 realize an effective optomechanical-type interaction \cite{Tsang2010}.
Such a system could convert states from the microwave to the optical domain
by driving sideband cooling transitions \cite{Marquardt2007, Wilson-Rae2007, Schliesser2008}.
Similar to optomechanical systems, 
the interaction requires
 large vacuum coupling rates and  the resolved-sideband regime \cite{Marquardt2007, Wilson-Rae2007, Schliesser2008}
 to be efficient as well as a optical cavity decay rate that greatly exceeds the microwave decay rate. Despite interest in the scheme, to date, it has not been realized. Previous demonstrations attained vacuum coupling rates of $\sim 2\pi \, \mathcal{O}(1-10)$Hz insufficient for an efficient transfer. In addition, several previous schemes operated with a microwave dissipation that was larger than the optical one, preventing efficient transfer from optical to microwave fields.
Here, we show that these two requirements can
be fulfilled in principle by coupling the
electric field of a superconducting resonator 
 to a whispering-gallery-mode (WGM) microresonator made from an electro-optical material, benefiting from the large vacuum electric field of superconducting circuits, akin to cQED \cite{Wallraff2004}.

\begin{figure}[tb]
	\centering
	\includegraphics[width=\columnwidth]{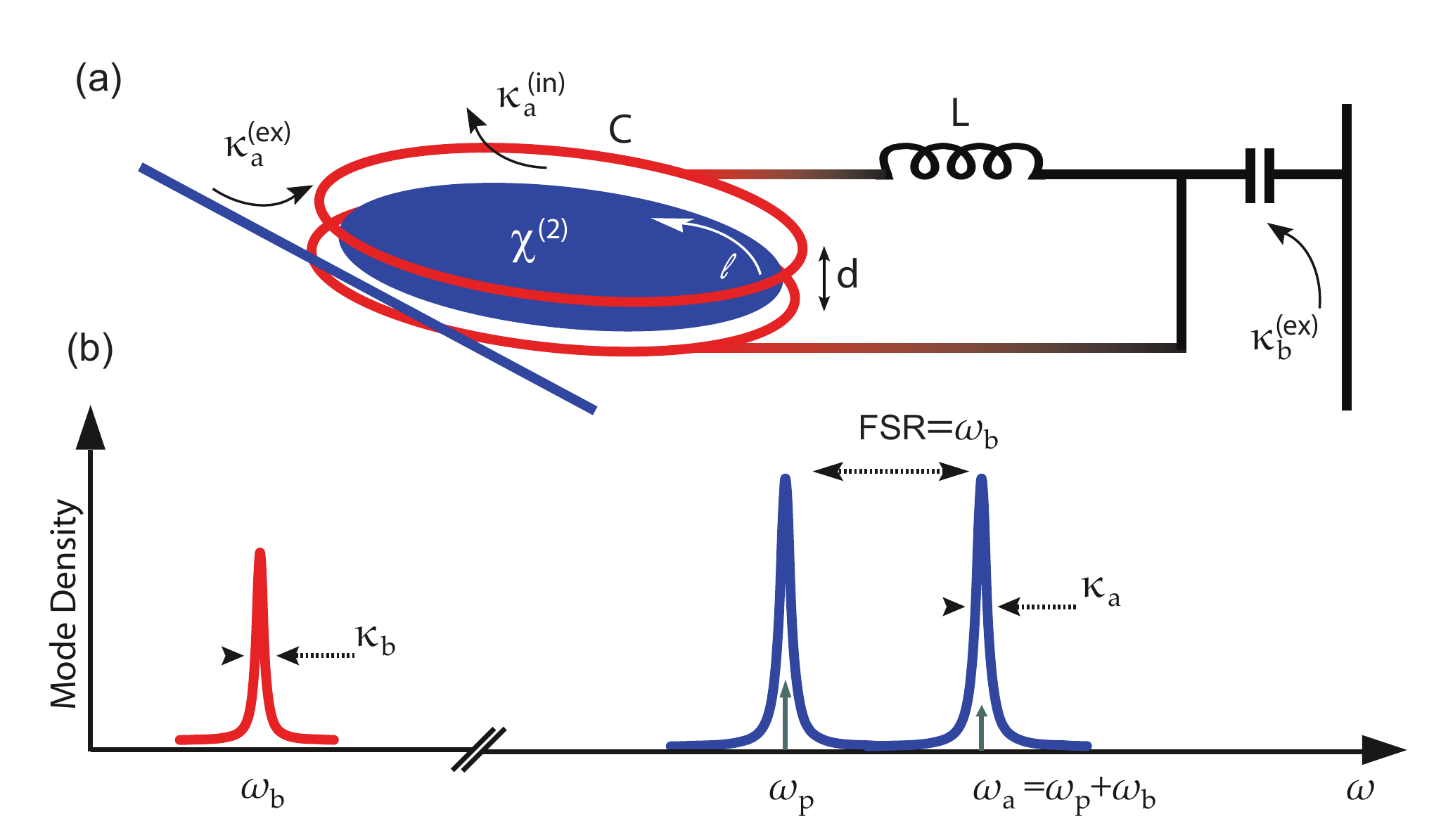}
	\caption{\label{fig:principle}
		(Color online) Principle for a cavity electro-optical system for microwave-to-optical conversion, constituted of (a) a $\chi^{(2)}$ WGM microresonator (in blue) resonant at $ \omega_{a} $ coupled to a waveguide with an external coupling rate $\kappa^{(ex)}_a$ and of intrinsic loss rate $\kappa^{(in)}_a$; and an LC microwave circuit, resonant at $ \omega_{b} $ coupled via a feedline (here capacitively coupled) with an external coupling rate $\kappa^{(ex)}_b$. $ d $ is the distance between the microwave resonator electrodes and $ l $ is the length of the optical path. (b) In the resolved sideband limit, and for the dissipation hierarchy $\kappa_a \gg \kappa_b$, the system enables via lower sideband pumping at $\omega_{\rm p} = \omega_a - \omega_b$ to laser cool the microwave photons, i.e. up-convert them to the cold optical mode, establishing a coherent interface between microwave and optical fields.
To minimize required optical input power it is possible to match the microwave resonance frequency $ \omega_{b} $ to the optical free spectral range (FSR).}
\end{figure}

\emph{Theoretical description}.
In the most general form, the cavity electro-optical dynamics is described 
by the Hamiltonian \cite{Tsang2010}
\begin{equation}
\label{eq:Hamiltonian}
\hat{H} = 
\hbar \omega_{a} \hat{a}^{\dagger}\hat{a} 
+ \hbar \omega_{b} \hat{b}^{\dagger}\hat{b} 
- \hbar g_0 \left( \hat{b} + \hat{b}^{\dagger} \right) 
\hat{a}^{\dagger}\hat{a},
\end{equation}
where 
$ a $ and 
$ a^{\dagger} $ ($ b $ and $ b^{\dagger} $) 
are the optical (microwave) annihilation and creation operators and 
$ \omega_{a} $ ($ \omega_{b} $) 
the optical (microwave) angular resonant frequency. 
The electro-optic coupling coefficient is given by

$g_0 = 
\tfrac{\omega_{a} n^{3} r l}{c \tau D} 
\sqrt{\tfrac{\hbar \omega_{b}}{2 C}}$
for a generic geometry (as depicted in Fig. \ref{fig:principle}), where $ n $ is the optical refractive index of the $ \chi^{(2)} $ nonlinear medium, $ r $ is its electro-optic coefficient, $ l $ is the length of the medium along the optical path (such that $ l=2 \pi r $), $ D $ is its thickness, and $ \tau $ is the optical round-trip time. $ \tfrac{1}{D}\sqrt{\tfrac{\hbar \omega_{b}}{2 C}}  = \tfrac{V_{\rm zpf}}{D} = \lvert \overrightarrow{E}_{\rm zpf}\rvert$ is the zero-point fluctuation of the microwave field at resonant frequency  $ \omega_{b} $.

The converter operation uses the fact that the electro-optical interaction is formally equivalent to the optomechanical Hamiltonian, whereby the microwave field plays the role of the mechanical degree of freedom. Consequently, in the good cavity limit (resolved sideband regime), i.e. $\omega_b\gg\kappa_a$, and when optical dissipation dominates, i.e. $\kappa_a \gg \kappa_b$, pumping the system with an optical laser on the lower sideband ($\omega_p=\omega_a -\omega_b$), will in the linearized regime lead to a beam-splitter interaction Hamiltonian $\hat{H}= \hbar g_0 \left( \hat{a}\hat{b}^{\dagger} + \hat{a}^{\dagger}\hat{b} \right)$ which effectively sideband cools the microwave mode, i.e. converts the microwave state to an optical photon at frequency $\omega_p+\omega_b$. For the case of a zero temperature bath and a pulsed optical cooling field, the input state of the microwave field and the optical field are swapped and state transfer achieved \cite{Tsang2010}. 
While electro-optical materials have been widely employed in modern optical telecommunication, realizing the conversion scheme in this manner has been challenging \cite{Ilchenko2003,Strekalov2009}, due to the inability to achieve large overlap of the microwave and optical field, resulting in insufficient coupling rates. Moreover, while crystalline WGM \cite{Ilchenko2004} can attain ultra high quality factors (Q), it poses stringent conditions on the microwave Q to achieve the proper dissipation hierarchy $\kappa_a\gg\kappa_b$ for conversion.

For a WGM resonator coupled to an LC circuit the electro-optic coupling coefficient can be expressed as
\begin{equation}
  \label{eq:Coupling2}
  g_0 
  = 
  \omega_{a} n^{2} r  
  \sqrt{\dfrac{\hbar \omega_{b}}
  {\varepsilon_{0} \varepsilon \mathcal{V}_{b}}},
\end{equation}
where the only geometrical parameter is the microwave mode volume $ \mathcal{V}_{b} $. 
Therefore to attain a large vacuum coupling rate $ g_0 $, 
a large overlap of the electric field distribution and the optical mode of the cavity has to be attained.
This formula also emphasizes that a material with high electro-optic coefficient $r$, high refractive index $n$ and low microwave dielectric constant $ \varepsilon $ are preferable.

\begin{figure}[h]
\centering
\includegraphics[width=.8\columnwidth]{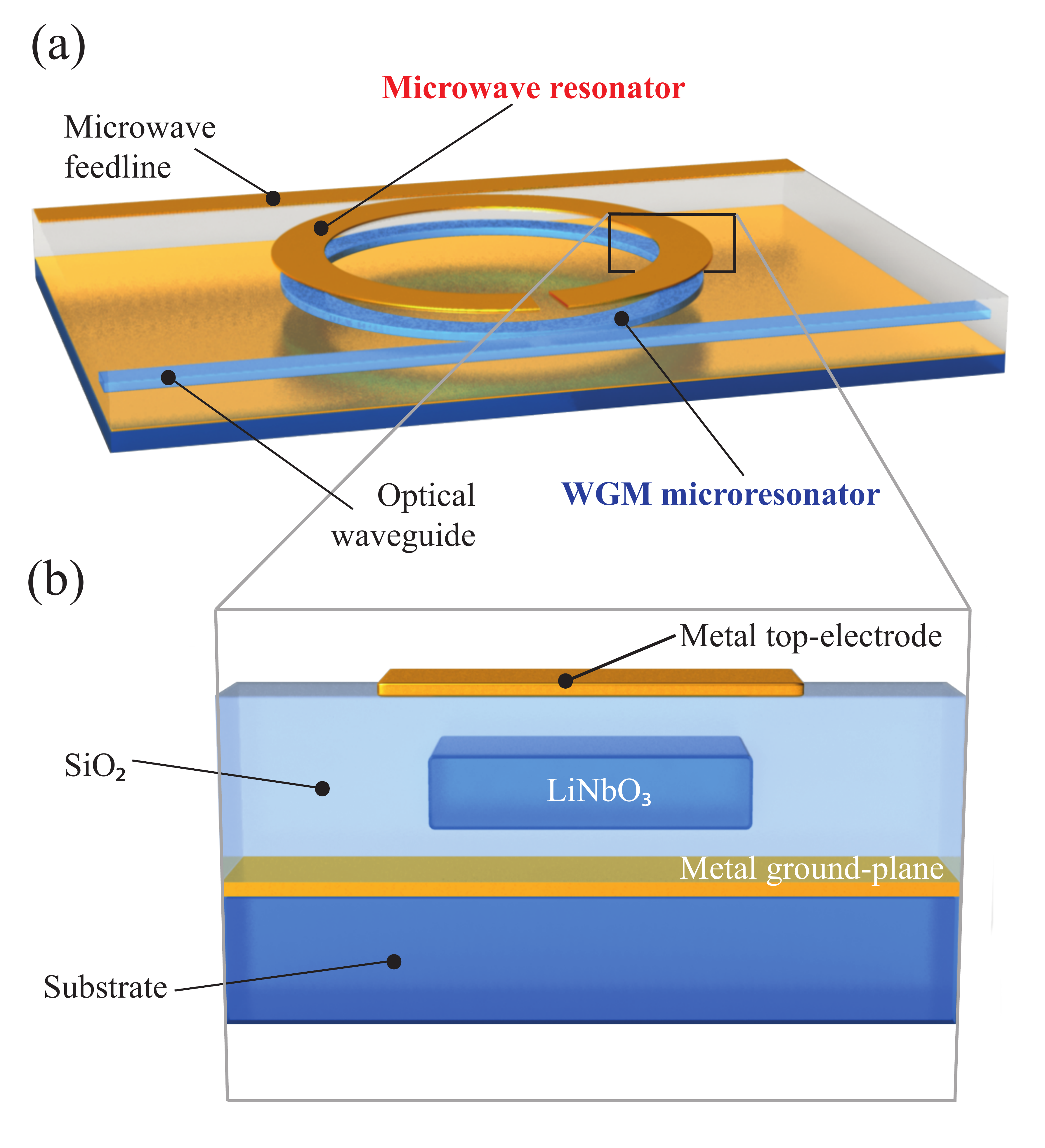}
\caption{\label{fig:device}
(Color online) Schematic view of the cavity quantum electro-optical system proposed for up-converting microwave photons to the optical domain. (a) depicts the top view and (b) shows a sample cross-section. The optical cavity is a microring resonator made out of z-cut $\mathrm{LiNbO}_{3} $ coupled via a waveguide by evanescent-field coupling. The microwave cavity is an open $ \tfrac{\lambda}{2} $ microstrip resonator. The resonator is overlapped with the optical microring to maximize the mode overlap and thus the electro-optic coupling. Note that the symmetry of the microwave resonator is broken, for example by grounding its ends, to ensure that only the positive phase of the microwave electric field profile couples to the optical microresonator. The microwave cavity is capacitively coupled to a microstrip feedline.}
\end{figure}

\emph{Device implementation}.
To enable sizeable electro-optical coupling to an integrated  nonlinear optical microresonator on the same chip, 
the proposed new underlying hybrid device architecture uses the large vacuum electric field offered by superconducting resonators, 
which confine electromagnetic modes to small volume $\mathcal{V}_{b} \ll \lambda^3$ 
and commonly exhibit high-Q. 
Indeed, titanium nitride TiN resonators can attain quality factors as high as 
$ Q_{b} \sim 10^{7} $ 
\cite{Leduc2010} at millikelvin temperatures.
The proposed on chip, integrated device is based on an optical WGM microresonator made from  a material that features $ \chi^{(2)} $ nonlinearity, 
such as lithium niobate ($\mathrm{LiNbO}_{3} $) or aluminium-nitride ($ \mathrm{Al}\mathrm{N}$) \cite{Guarino2007, Xiong2012}. 
As shown schematically in Fig.~\ref{fig:device}, 
the planar microresonator is coupled to an open superconducting microstrip resonator, 
whose electric  field couples to the optical mode via the electro-optical effect. Note here that the symmetry of the microwave resonator must be broken to ensure that only the positive phase of the microwave electric field profile couples to the optical microresonator.
The fabrication of microresonators from electro-optical materials is made possible via crystalline 
$\mathrm{LiNbO}_{3} $ thin films
\footnote{which has recently become commercially available from NanoLN}, which allow to combine the large on-chip element density of integrated photonics with the second-order nonlinearity of $\mathrm{LiNbO}_{3} $. 
Microresonators with Q $\sim 10^6$ have been demonstrated with this material \cite{Wang2015}.

Because of the absence of  a symmetry center, 
nonlinear $ \chi^{(2)} $ materials 
also exhibit piezoelectricity, which can cause modulation of 
the optical field \cite{Wang2014a} and perturb the electro-optic modulation via the Pockels effect. 
By design, the 
$\mathrm{LiNbO}_{3}$ microring 
is embedded in silica
($ \mathrm{SiO}_{2} $) 
and thus is clamped. 
Hence, the mechanical degree of freedom is frozen and the piezoelectric contribution to the modulation made negligible. 
This result was verified by a simulation comparing a suspended microdisk and an embedded microring of same geometries under the same microwave excitation. 
For the latter, the piezoelectric coupling strength is more than 9 orders of magnitude smaller and therefore can be neglected. 
Our device and simulation efforts focused on Z-cut 
$\mathrm{LiNbO}_{3} $ 
after having discarded other potential nonlinear materials. 
$\mathrm{LiNbO}_{3} $ 
exhibits a $ r_{51} $ as high as 
30$\,\mathrm{pm \cdot V^{-1}} $. 
Nanofabrication platforms are also mature enough to provide good structures with thin-film single crystals 
\cite{Poberaj2012, Xiong2012}. 
We conducted numerical simulations in order to take into account the anisotropy of the material and the complex geometry of our design.

\emph{Numerical simulations}.
\begin{figure}[t]
\centering
\includegraphics[width=1\columnwidth]{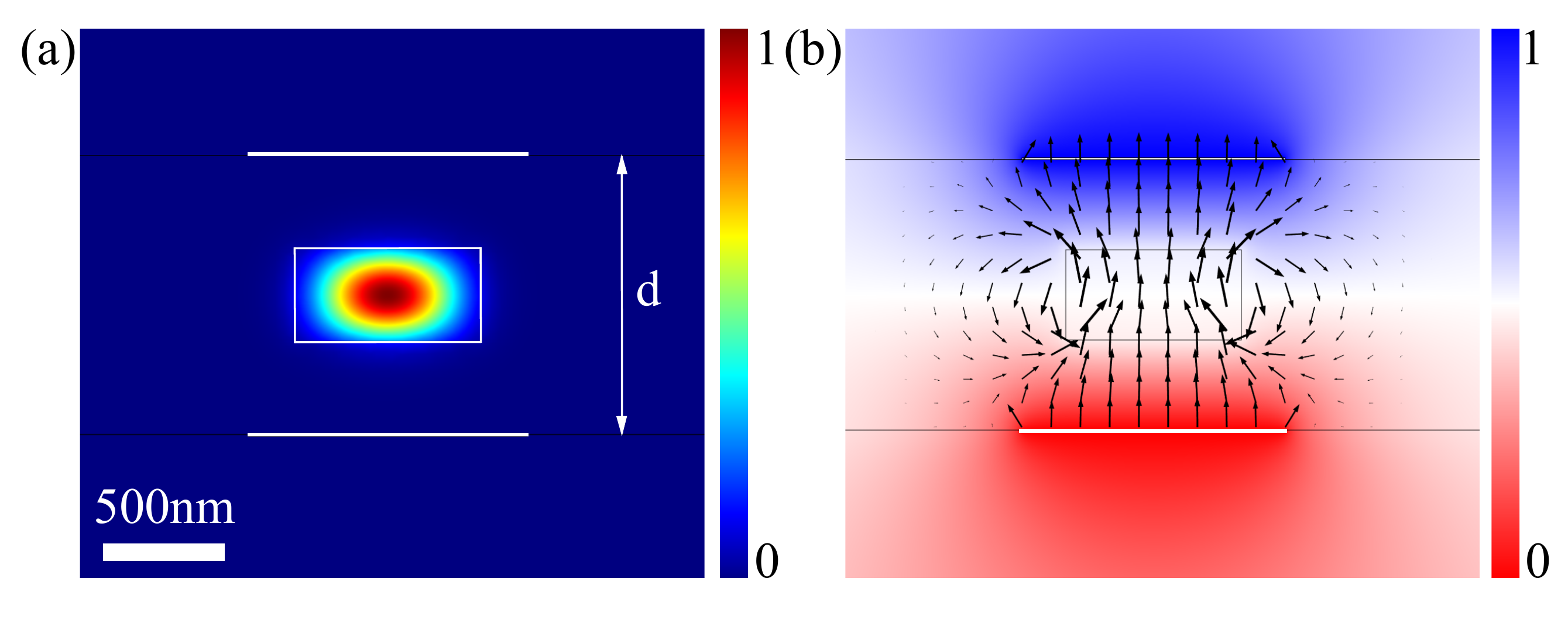}
\caption{\label{fig:FEM}(Color online) 2D FEM simulations of an optimized geometry. (a) Optical energy density profile of an optical WGM mode in a lithium niobate resonator with $ \omega_{a} = 2\pi \cdot 200\,{\rm THz}$. The refractive index contrast between the $\mathrm{LiNbO}_{3}$ WGM and the surrounding $\mathrm{SiO}_{2}$ provides high confinement of the optical energy. (b) Electric potential and electric field (arrows) across the electrodes inside the microwave cavity at $ \omega_{b} = 2\pi \cdot 6\,{\rm GHz}$. Direction and amplitude of the field are optimized to maximize $g_{0}$. $ d $ is the distance between the electrodes. Plots are on log scale with color map given on the right in arbitrary units.}
\end{figure}
In order to numerically evaluate the coupling coefficient of our proposed design we use finite element (FEM) simulation of the optical modes via the weak formulation of the 
Helmholtz equation \cite{Oxborrow2007}. 
We solve for eigenstates of the system in 2D representation with 
 an axial symmetry simplification \cite{Oxborrow2007}.
 A FEM simulation of the optical mode for LiNbO$_3$ is shown in
Fig.~\ref{fig:FEM}.
Since the WGM resonator has $ \chi^{(2)} $ 
nonlinearity, an external microwave field 
$ \textbf{E}_{b} = - \mathbf{\nabla} V$ 
will change the refractive index 
$ [n]= \left( \sqrt{ [\eta]  } \right)^{-1}$ and thus modify the optical field. 
To introduce the electro-optical interaction, we use an expansion of the impermeability
\begin{equation}
\label{eq:EO_effect}
  \eta ( \textbf{E}_{b}) _{ij} 
  = 
  \eta ( \textbf{0})_{ij} + r_{ijk} \cdot E_{b}^{k} 
  + \zeta_{ijkl} \cdot E_{b}^{k}  E_{b}^{l} 
  + \ldots
\end{equation}
We only keep the second term in the expansion corresponding to the second-order nonlinear effect, so called Pockels effect.
The electro-optic tensor being not axisymmetric, thus one has to integrate the electro-optical interaction over the angular coordinate along the WGM optical mode. 

To evaluate the (vacuum) electro-optical coupling rate 
$g_{0}/2\pi$, one has to compute the frequency shift 
$\delta \omega_{a}$ caused 
by the microwave zero-point voltage fluctuations  
$V_{\rm zpf}$ 
and corresponding 
$ \textbf{E}_{b} $. 
Then the electro-optic coefficient reads
$ 
  g_0 
  = 
  V_{\rm zpf} \cdot 
  \tfrac{\delta \omega_{a}}{V}
$
where $ V $ is the voltage applied for the simulation and $V_{\rm zpf}$ depends only on the microwave field distribution that we simulate (as shown in
Fig.~\ref{fig:FEM}). 

The Bethe-Schwinger cavity perturbation 
formula \cite{Schwinger1943} 
can be used at a first-order approximation assuming that the perturbation $ \delta \varepsilon $ has a small effect on the cavity:
\begin{equation}
\label{eq:Bethe}
 \dfrac{\delta \omega_{a}}{\omega_{a}} 
 \approx 
 \dfrac{\int_{\mathcal{V}} \textbf{E}_{a}^{*}\varepsilon_{0} 
 \delta \varepsilon\left( \textbf{r} \right) 
 \textbf{E}_{a} d\mathcal{V}}{\int_{\mathcal{V}} 
 \varepsilon_{0} \varepsilon\left( \textbf{r} \right) 
 \textbf{E}_{a}^{*} \cdot \textbf{E}_{a}  d\mathcal{V}},
\end{equation}
where 
$ \mathcal{V} $ is the integration volume. 
For a high-Q resonator the radiation losses are small and the integration volume can be taken over the boundaries of the resonator. 
Eq. (\ref{eq:Bethe}) can be written as
$
  \tfrac{\delta \omega_{a}}{\omega_{a}} 
  = 
  \tfrac{\delta U_{a}}{U_{a}}
$
i.e. as the ratio between the optical energy introduced by the perturbation 
($ \delta U_{a} $) 
and the total energy of the unperturbed cavity 
($ U_{a} $).
As the effect is very weak, the optical energy variation is induced only by a modification of the resonator 
permeability constant $  \delta \varepsilon_{ij} 
  = 
  \varepsilon_{ik}\cdot \varepsilon_{jl}\cdot \delta \eta_{kl}.$
One can write
\begin{equation}
\label{eq:deltaU}
  \delta U_{a} 
  = 
  \dfrac{1}{2} \varepsilon_0 \int_{\mathcal{V}} 
  \delta \varepsilon_{ij}\ E^{i}_a E^{j}_a d\mathcal{V} 
  =
  \dfrac{1}{2} \varepsilon_0 \int_{\mathcal{V}} 
  \delta \eta_{kl}\ D^{k}_a D^{l}_a d\mathcal{V},
\end{equation}
where 
$D^{k}_a$ is an unperturbed electric displacement field of an optical mode. 
$ g_0 $ can then be determined by separately simulating the optical and microwave fields' distributions.

Therefore, taking into account the geometry and the anisotropy of the system, the expression (\ref{eq:Coupling2}) of the electro-optical coupling coefficient  becomes
\begin{equation}
\label{eq:g0_final}
  g_{0} 
  =  
  \dfrac{\omega_{a} \varepsilon_0}{2 U_{a} V} 
  \sqrt{\dfrac{\hbar \omega_{b}}{2 C}}  
  \int_{\mathcal{V}} \varepsilon_{ik}\cdot 
  \varepsilon_{jl}\cdot r_{klm} \cdot E^m_b E^{i}_a E^{j}_a d\mathcal{V}.
\end{equation}

An optimization of the different parameters, such as position and size of the microwave microstrip or polarization, was run in order to maximize $ g_{0} $. In particular, from Eq. (\ref{eq:deltaU}) one can extract that TE optical modes characterized by high axial components $ D^{l}_z $ give higher coupling. For the geometry in Fig.~\ref{fig:FEM}, we compute $ g_{0} \sim 2\pi\cdot50\,\mathrm{kHz}$ at $ \omega_{a}=2\pi \cdot 200\,\mathrm{THz} $ (i.e. $\lambda_a\approx1550$~nm) and $\omega_{b}=2\pi \cdot 6\,\mathrm{GHz} $ with $ d \sim 1.5\,\mathrm{\mu m} $.
This value is more than 3 orders of magnitude larger than previous work \cite{Ilchenko2003,Strekalov2009}.
The position of the electrodes is critical. Electrodes on top (as in Fig.~\ref{fig:FEM}) proves to give higher coupling and are more convenient for fabrication (compared to electrodes on the side of the optical WGM, with similar distance to the WGM). The distance $ d $ between the electrodes is key to provide high-confinement of the field. However it must be kept in mind as well that the quality factors of both optical and microwave cavities depend on the geometry. For instance, no metal should directly be too close to the evanescent field of the optical waveguide fields. For the geometries G1-G4 in Table~\ref{table}, the field energy density at the position of the electrodes is 6 orders of magnitude smaller than that at the center of the mode, and therefore absorption is strongly reduced. 
Parameters and results are detailed in Table \ref{table} for different optimum geometries.

\begin{table}
\caption{\label{table}Parameters and results for different geometries in LiNbO$_{3} $. G1: electrodes on the side of WGM (optimized distance $ d \sim 1.5\,\mathrm{\mu m} $, as in Fig.~\ref{fig:FEM}). G2: electrodes on top and bottom of WGM (not optimized distance). G3: electrodes on top and bottom of WGM (optimized distance). G4: electrodes on top and bottom of WGM (optimized distance and polarization), but axial polarization of $ \textbf{E}_{a} $ (instead of radial). Parameters are computed for realistic experimental parameters: $ \tfrac{\omega_{a}}{2\pi}=200\,\mathrm{THz} $, $ \tfrac{\omega_{b}}{2\pi}=6\,\mathrm{GHz} $, $ Q_{a} \simeq 10^{5} $ and $ Q_{b} \simeq 10^{3} $.}
\begin{ruledtabular}
\begin{tabular}{l|cccc}
Geometry & G1 & G2 & G3 & G4\\
\hline
Model & \includegraphics[scale=.15]{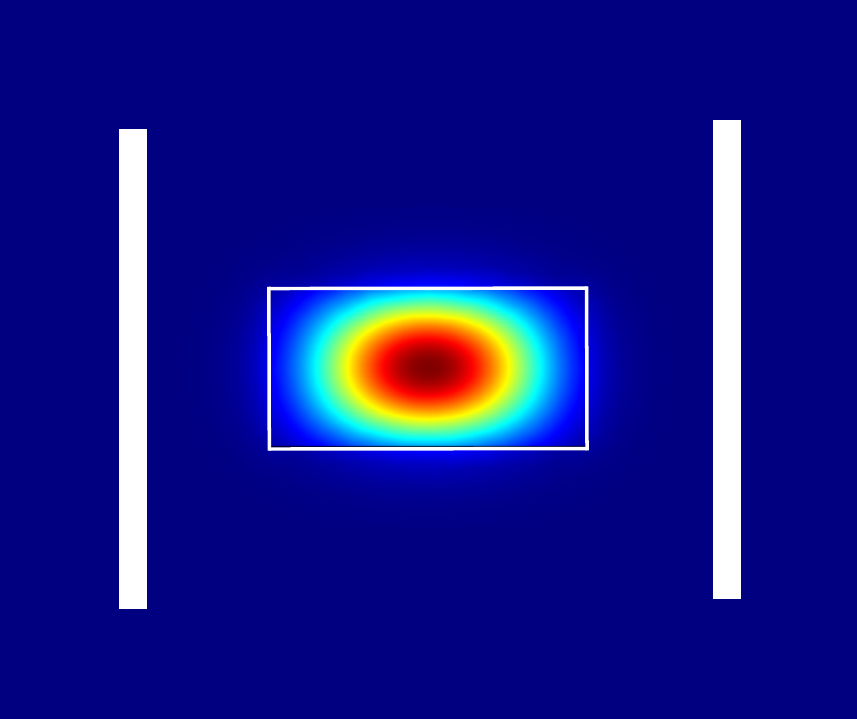}  &
\includegraphics[scale=.15]{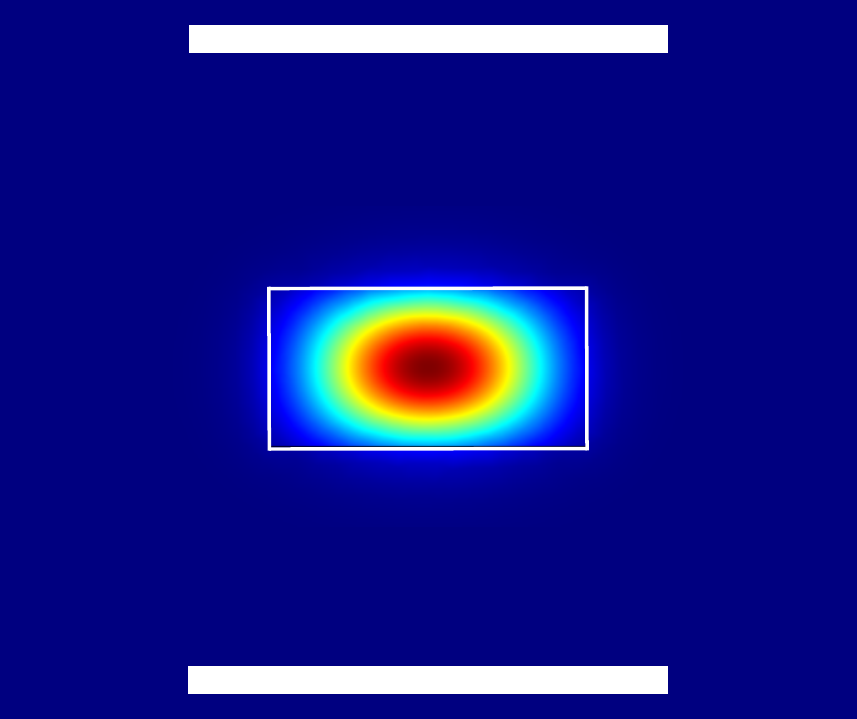} &
\includegraphics[scale=.15]{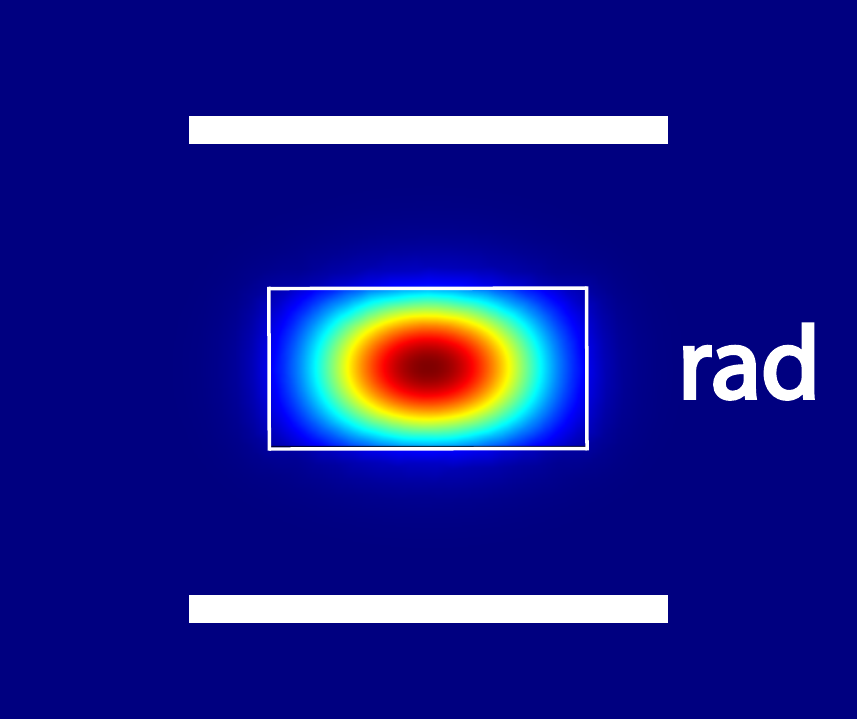} &
\includegraphics[scale=.15]{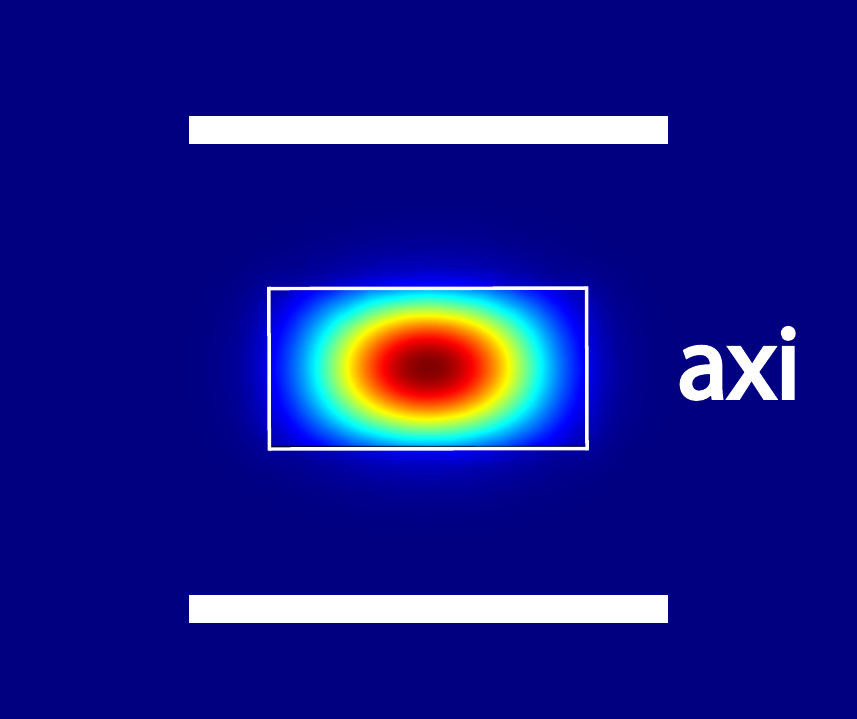}\\
\hline
$ {g_{0}}/{2\pi} $ (kHz) & 0.15 & 0.75 & 12 & 50\\
\hline
$ \mathcal{C}_{0}={4 \lvert g_{0} \rvert^{2}}/\left(\kappa_{a}\kappa_{b}\right) $ & $8\cdot 10^{-12}$ & $2\cdot 10^{-10}$ & $5\cdot 10^{-8}$ & $9\cdot 10^{-7}$\\
\hline
$ P $ (W) {\scriptsize -- single mode}  & $7500$ & $300$ & $1.2$ & $0.067$\\
\hline
$ P $ (W) {\scriptsize -- dual mode} & $200$ & $8$ & $0.03$ & $0.0018$\\
\end{tabular}
\end{ruledtabular}
\label{}
\end{table}

\emph{Scheme}.
By pumping the lower sideband of the hybrid electro\-optical system at 
$\omega_{\rm p} = \omega_a - \omega_b$ 
(see Fig.~\ref{fig:principle}) the microwave photons are laser cooled, i.e. 
up-converted to the cold optical mode. 
If the microwave cavity is already cooled to the ground state by passive cooling, 
which is easily achieved in the case of superconducting cavity with GHz resonance frequency cooled to the base temperature of a dilution refrigerator, 
this up-conversion may be used to establish a coherent interface between microwave and optical fields. 
The proposed converter would work at  
$ \omega_{b} = 2\pi \cdot 6\, \mathrm{GHz} $, 
a typical frequency of microwave superconducting qubits~\cite{Barends2013}, 
and 10~mK to assure the microwave cavity ground state.

The main objective of the device is to achieve quantum coherent microwave-to-optical frequency conversion:
to achieve near-complete frequency conversion (i.e. quantum efficiency $\gamma\sim 1$),  the extrinsic decay rate $ \kappa^{(ex)}_a $ of the optical cavity should dominate the intrinsic one $ \kappa^{(in)}_a $ and the effective coupling rate between the optical and the microwave modes should satisfy the hierarchy 
$\kappa^{(ex)}_a \gg\{2g_0 \sqrt{\bar{n}_p},\kappa^{(in)}_a\} $,  where $\bar{n}_p$ 
is the number of photons in the optical cavity, 
such that the frequency up-converted microwave photon leaves the optical cavity before 
it decays or is down-converted back to the microwave frequency range; 
the microwave cavity should be strongly overcoupled ($\kappa^{(ex)}_b \gg \kappa^{(in)}_b$) .
The overall intrinsic efficiency of the frequency up-conversion process  $\gamma$, defined by the efficiency of converting microwave to optical quanta,
can be calculated as
\begin{equation}
\label{eq:efficiency}
  \gamma \left[ \omega \right]  =
  \dfrac{\kaex}{\kappa_a}
  \dfrac{\kbex}{\kappa_b}
  \dfrac{ 4 \mathcal{C}}{\left( 1 + \mathcal{C} \right)^{2}}
  \dfrac{1}{1 + \dfrac{\left( \omega_b-\omega \right)^{2}}
  {\kappa_b^{2} \left( 1+\mathcal{C} \right)^{2} /4}};
\end{equation}
thus the overall efficiency is a function of frequency and cooperativity $ \mathcal{C} $, which is proportional to
the number of photons in the optical cavity 
$\mathcal{C} = \bar{n}_p \cdot \mathcal{C}_{0}$, 
where $\mathcal{C}_{0}=\tfrac{4 \lvert g_{0} \rvert^{2}}{\kappa_{a}\kappa_{b}}$ 
is the electrooptical single-photon cooperativity. 
Interestingly, in our proposed scheme when $ \omega=\omega_b $, 
a cooperativity of $ \mathcal{C}=1 $ is sufficient to obtain the maximum efficiency~\cite{Tsang2011}, 
making achievable the complete photon conversion between microwave and optical fields on a chip.

The quantum-state-transfer is only possible if a unity cooperativity  is achieved 
and the coupling strength is much larger than the microwave decay rates 
$ \kappa_{b} $, i.e. $2g_0 \sqrt{\bar{n}_p}\gg\kappa_b$.
When implementing the converter with a single optical resonance (cf. Table~\ref{table}) 
the required power levels are of the order of $100$ milliwatts or even higher, making a realistic implementation impossible. 
The power can however be substantially reduced by employing two optical modes (cf. Fig.~\ref{fig:principle}), 
spaced by the microwave frequency (i.e. a multiple cavity mode transducer ~\cite{Dobrindt2010}). 
In this case, when $  \Delta \omega_{a} = \omega_{b} $, one has
\begin{equation}
  \bar{n}_p 
  = 
  \dfrac{P}{\hbar \omega_{a} \kappa_a},
\end{equation}
where $ P $ is the optical pump power.
With such a dual mode design, the pump power required to obtain
$ \mathcal{C} = 1 $ is $ P = \hbar \omega_{a} \kappa_a\mathcal{C}^{-1}_{0} $ 
and can be as low as $ \mathcal{O}(1)\,\mathrm{mW} $ 
with conservative parameters slightly better than 
$ Q_{a} \simeq 10^{5} $ and $ Q_{b} \simeq 10^{3} $. 
Thus according to expression~(\ref{eq:efficiency}), 
a conversion efficiency exceeding 90\% can be achieved with $\mathcal{O}(1)\,\mathrm{mW}$ of pump power with the presented design. 
In contrast to the realization with the only optical mode, 
in the dual-optical-mode scheme 
the required power does reduce when $Q_{a}$ increases. 
For instance, for the state-of-the-art parameters of 
$ Q_{a} \simeq 10^{6} $ and $ Q_{b} \simeq 10^{4} $ 
one would need $  \mathcal{O}(1)\,\mathrm{\mu W} $ of optical power only.

The noise introduced by the coupling of modes other than the incoming microwave signal limits the quantum fidelity of the microwave to optical photon conversion.
We theoretically characterized the noise added during the conversion process by the equivalent quanta of the total noise, as compared to the spectral density of the input signal.
On resonance, we find that
\begin{equation}
  n_\mathrm{eq} =
  \dfrac{\kappa_{b}}{\kappa^{(ex)}_{b}}\left(
  2 \nelth
  +
  \dfrac{\left( 1+\mathcal{C} \right)^2}{4 \mathcal{C}}
  \dfrac{\kappa_a}{\kappa^{(ex)}_{a}}
  \right)
\end{equation}
quanta of noise are introduced,
with $ \nelth$ the thermal occupation of the microwave mode.
The first term gives the noise contribution from the microwave, and the second from the optical degrees of freedom. For strongly overcoupled resonators, one would only add $ n_\mathrm{eq} = 2 \nelth+1 $.

\emph{Conclusions}.
In conclusion, we propose a direct quantum electrooptical converter based on a 
$ \chi^{(2)} $ nonlinear WGM microresonator coupled to a planar superconducting microwave resonator which realizes large coupling rates. 
Electro-optical coupling coefficient as high as 
$g_0$ $\sim 2\pi \, \mathcal{O}(10-100)$~kHz  can be attained with 
LiNbO$_{3} $ thin films. 
We show that high-fidelity conversion can be achieved based on current technology with optical power as low as 
$ \mathcal{O}(1)\,\mathrm{mW} $. 
On-chip, it will become efficiently coupled to and controlled by optical fields and thus may enable new regimes for radio- and microwave electromagnetic field detection 
(allowing quantum-limited amplification and readout of microwave and radio-frequency radiation). 
For instance, the proposed device could enable quantum microwave illumination \cite{Pirandola2015}. 
\begin{acknowledgments}
This work was financially supported by the SNF, 
the NCCR Quantum Science and Technology 
and the European Union Seventh Framework Program 
through iQUOEMS (grant no.323924). 
TJK acknowledges financial support from an ERC AdG (QREM).
\end{acknowledgments}

\bibliography{library}

\begin{thebibliography}{34}%
\makeatletter
\providecommand \@ifxundefined [1]{%
 \@ifx{#1\undefined}
}%
\providecommand \@ifnum [1]{%
 \ifnum #1\expandafter \@firstoftwo
 \else \expandafter \@secondoftwo
 \fi
}%
\providecommand \@ifx [1]{%
 \ifx #1\expandafter \@firstoftwo
 \else \expandafter \@secondoftwo
 \fi
}%
\providecommand \natexlab [1]{#1}%
\providecommand \enquote  [1]{``#1''}%
\providecommand \bibnamefont  [1]{#1}%
\providecommand \bibfnamefont [1]{#1}%
\providecommand \citenamefont [1]{#1}%
\providecommand \href@noop [0]{\@secondoftwo}%
\providecommand \href [0]{\begingroup \@sanitize@url \@href}%
\providecommand \@href[1]{\@@startlink{#1}\@@href}%
\providecommand \@@href[1]{\endgroup#1\@@endlink}%
\providecommand \@sanitize@url [0]{\catcode `\\12\catcode `\$12\catcode
  `\&12\catcode `\#12\catcode `\^12\catcode `\_12\catcode `\%12\relax}%
\providecommand \@@startlink[1]{}%
\providecommand \@@endlink[0]{}%
\providecommand \url  [0]{\begingroup\@sanitize@url \@url }%
\providecommand \@url [1]{\endgroup\@href {#1}{\urlprefix }}%
\providecommand \urlprefix  [0]{URL }%
\providecommand \Eprint [0]{\href }%
\providecommand \doibase [0]{http://dx.doi.org/}%
\providecommand \selectlanguage [0]{\@gobble}%
\providecommand \bibinfo  [0]{\@secondoftwo}%
\providecommand \bibfield  [0]{\@secondoftwo}%
\providecommand \translation [1]{[#1]}%
\providecommand \BibitemOpen [0]{}%
\providecommand \bibitemStop [0]{}%
\providecommand \bibitemNoStop [0]{.\EOS\space}%
\providecommand \EOS [0]{\spacefactor3000\relax}%
\providecommand \BibitemShut  [1]{\csname bibitem#1\endcsname}%
\let\auto@bib@innerbib\@empty
\bibitem [{\citenamefont {Hofheinz}\ \emph {et~al.}(2009)\citenamefont
  {Hofheinz}, \citenamefont {Wang}, \citenamefont {Ansmann}, \citenamefont
  {Bialczak}, \citenamefont {Lucero}, \citenamefont {Neeley}, \citenamefont
  {O'Connell}, \citenamefont {Sank}, \citenamefont {Wenner}, \citenamefont
  {Martinis},\ and\ \citenamefont {Cleland}}]{Hofheinz2009}%
  \BibitemOpen
  \bibfield  {author} {\bibinfo {author} {\bibfnamefont {M.}~\bibnamefont
  {Hofheinz}}, \bibinfo {author} {\bibfnamefont {H.}~\bibnamefont {Wang}},
  \bibinfo {author} {\bibfnamefont {M.}~\bibnamefont {Ansmann}}, \bibinfo
  {author} {\bibfnamefont {R.~C.}\ \bibnamefont {Bialczak}}, \bibinfo {author}
  {\bibfnamefont {E.}~\bibnamefont {Lucero}}, \bibinfo {author} {\bibfnamefont
  {M.}~\bibnamefont {Neeley}}, \bibinfo {author} {\bibfnamefont {a.~D.}\
  \bibnamefont {O'Connell}}, \bibinfo {author} {\bibfnamefont {D.}~\bibnamefont
  {Sank}}, \bibinfo {author} {\bibfnamefont {J.}~\bibnamefont {Wenner}},
  \bibinfo {author} {\bibfnamefont {J.~M.}\ \bibnamefont {Martinis}}, \ and\
  \bibinfo {author} {\bibfnamefont {a.~N.}\ \bibnamefont {Cleland}},\ }\href
  {\doibase 10.1038/nature08005} {\bibfield  {journal} {\bibinfo  {journal}
  {Nature}\ }\textbf {\bibinfo {volume} {459}},\ \bibinfo {pages} {546}
  (\bibinfo {year} {2009})}\BibitemShut {NoStop}%
\bibitem [{\citenamefont {Devoret}\ and\ \citenamefont
  {Schoelkopf}(2013)}]{Devoret2013}%
  \BibitemOpen
  \bibfield  {author} {\bibinfo {author} {\bibfnamefont {M.~H.}\ \bibnamefont
  {Devoret}}\ and\ \bibinfo {author} {\bibfnamefont {R.~J.}\ \bibnamefont
  {Schoelkopf}},\ }\href {\doibase 10.1126/science.1231930} {\bibfield
  {journal} {\bibinfo  {journal} {Science}\ }\textbf {\bibinfo {volume}
  {339}},\ \bibinfo {pages} {1169} (\bibinfo {year} {2013})},\ \Eprint
  {http://arxiv.org/abs/http://science.sciencemag.org/content/339/6124/1169.full.pdf}
  {http://science.sciencemag.org/content/339/6124/1169.full.pdf} \BibitemShut
  {NoStop}%
\bibitem [{\citenamefont {Barzanjeh}\ \emph {et~al.}(2012)\citenamefont
  {Barzanjeh}, \citenamefont {Abdi}, \citenamefont {Milburn}, \citenamefont
  {Tombesi},\ and\ \citenamefont {Vitali}}]{Barzanjeh2012}%
  \BibitemOpen
  \bibfield  {author} {\bibinfo {author} {\bibfnamefont {S.}~\bibnamefont
  {Barzanjeh}}, \bibinfo {author} {\bibfnamefont {M.}~\bibnamefont {Abdi}},
  \bibinfo {author} {\bibfnamefont {G.~J.}\ \bibnamefont {Milburn}}, \bibinfo
  {author} {\bibfnamefont {P.}~\bibnamefont {Tombesi}}, \ and\ \bibinfo
  {author} {\bibfnamefont {D.}~\bibnamefont {Vitali}},\ }\href {\doibase
  10.1103/PhysRevLett.109.130503} {\bibfield  {journal} {\bibinfo  {journal}
  {Physical Review Letters}\ }\textbf {\bibinfo {volume} {109}},\ \bibinfo
  {pages} {130503} (\bibinfo {year} {2012})}\BibitemShut {NoStop}%
\bibitem [{\citenamefont {Kimble}(2008)}]{Kimble2008}%
  \BibitemOpen
  \bibfield  {author} {\bibinfo {author} {\bibfnamefont {H.~J.}\ \bibnamefont
  {Kimble}},\ }\href {\doibase 10.1038/nature07127} {\bibfield  {journal}
  {\bibinfo  {journal} {Nature}\ }\textbf {\bibinfo {volume} {453}},\ \bibinfo
  {pages} {1023} (\bibinfo {year} {2008})}\BibitemShut {NoStop}%
\bibitem [{\citenamefont {Tian}(2015)}]{Tian2015}%
  \BibitemOpen
  \bibfield  {author} {\bibinfo {author} {\bibfnamefont {L.}~\bibnamefont
  {Tian}},\ }\href {\doibase 10.1002/andp.201400116} {\bibfield  {journal}
  {\bibinfo  {journal} {Annalen der Physik}\ }\textbf {\bibinfo {volume}
  {527}},\ \bibinfo {pages} {1} (\bibinfo {year} {2015})}\BibitemShut {NoStop}%
\bibitem [{\citenamefont {Schoelkopf}\ and\ \citenamefont
  {Girvin}(2008)}]{Schoelkopf2008}%
  \BibitemOpen
  \bibfield  {author} {\bibinfo {author} {\bibfnamefont {R.~J.}\ \bibnamefont
  {Schoelkopf}}\ and\ \bibinfo {author} {\bibfnamefont {S.~M.}\ \bibnamefont
  {Girvin}},\ }\href {\doibase 10.1038/451664a} {\bibfield  {journal} {\bibinfo
   {journal} {Nature}\ }\textbf {\bibinfo {volume} {451}},\ \bibinfo {pages}
  {664} (\bibinfo {year} {2008})}\BibitemShut {NoStop}%
\bibitem [{\citenamefont {Bochmann}\ \emph {et~al.}(2013)\citenamefont
  {Bochmann}, \citenamefont {Vainsencher}, \citenamefont {Awschalom},\ and\
  \citenamefont {Cleland}}]{Bochmann2013}%
  \BibitemOpen
  \bibfield  {author} {\bibinfo {author} {\bibfnamefont {J.}~\bibnamefont
  {Bochmann}}, \bibinfo {author} {\bibfnamefont {A.}~\bibnamefont
  {Vainsencher}}, \bibinfo {author} {\bibfnamefont {D.~D.}\ \bibnamefont
  {Awschalom}}, \ and\ \bibinfo {author} {\bibfnamefont {A.~N.}\ \bibnamefont
  {Cleland}},\ }\href {\doibase 10.1038/nphys2748} {\bibfield  {journal}
  {\bibinfo  {journal} {Nature Physics}\ }\textbf {\bibinfo {volume} {9}},\
  \bibinfo {pages} {712} (\bibinfo {year} {2013})}\BibitemShut {NoStop}%
\bibitem [{\citenamefont {Bagci}\ \emph {et~al.}(2014)\citenamefont {Bagci},
  \citenamefont {Simonsen}, \citenamefont {Schmid}, \citenamefont {Villanueva},
  \citenamefont {Zeuthen}, \citenamefont {Appel}, \citenamefont {Taylor},
  \citenamefont {Sorensen}, \citenamefont {Usami}, \citenamefont {Schliesser},\
  and\ \citenamefont {Polzik}}]{Bagci2014}%
  \BibitemOpen
  \bibfield  {author} {\bibinfo {author} {\bibfnamefont {T.}~\bibnamefont
  {Bagci}}, \bibinfo {author} {\bibfnamefont {A.}~\bibnamefont {Simonsen}},
  \bibinfo {author} {\bibfnamefont {S.}~\bibnamefont {Schmid}}, \bibinfo
  {author} {\bibfnamefont {L.~G.}\ \bibnamefont {Villanueva}}, \bibinfo
  {author} {\bibfnamefont {E.}~\bibnamefont {Zeuthen}}, \bibinfo {author}
  {\bibfnamefont {J.}~\bibnamefont {Appel}}, \bibinfo {author} {\bibfnamefont
  {J.~M.}\ \bibnamefont {Taylor}}, \bibinfo {author} {\bibfnamefont
  {A.}~\bibnamefont {Sorensen}}, \bibinfo {author} {\bibfnamefont
  {K.}~\bibnamefont {Usami}}, \bibinfo {author} {\bibfnamefont
  {A.}~\bibnamefont {Schliesser}}, \ and\ \bibinfo {author} {\bibfnamefont
  {E.~S.}\ \bibnamefont {Polzik}},\ }\href {\doibase 10.1038/nature13029}
  {\bibfield  {journal} {\bibinfo  {journal} {Nature}\ }\textbf {\bibinfo
  {volume} {507}},\ \bibinfo {pages} {81} (\bibinfo {year} {2014})}\BibitemShut
  {NoStop}%
\bibitem [{\citenamefont {Andrews}\ \emph {et~al.}(2014)\citenamefont
  {Andrews}, \citenamefont {Peterson}, \citenamefont {Purdy}, \citenamefont
  {Cicak}, \citenamefont {Simmonds}, \citenamefont {Regal},\ and\ \citenamefont
  {Lehnert}}]{Andrews2014}%
  \BibitemOpen
  \bibfield  {author} {\bibinfo {author} {\bibfnamefont {R.~W.}\ \bibnamefont
  {Andrews}}, \bibinfo {author} {\bibfnamefont {R.~W.}\ \bibnamefont
  {Peterson}}, \bibinfo {author} {\bibfnamefont {T.~P.}\ \bibnamefont {Purdy}},
  \bibinfo {author} {\bibfnamefont {K.}~\bibnamefont {Cicak}}, \bibinfo
  {author} {\bibfnamefont {R.~W.}\ \bibnamefont {Simmonds}}, \bibinfo {author}
  {\bibfnamefont {C.~A.}\ \bibnamefont {Regal}}, \ and\ \bibinfo {author}
  {\bibfnamefont {K.~W.}\ \bibnamefont {Lehnert}},\ }\href {\doibase
  10.1038/nphys2911} {\bibfield  {journal} {\bibinfo  {journal} {Nature
  Physics}\ }\textbf {\bibinfo {volume} {10}},\ \bibinfo {pages} {321}
  (\bibinfo {year} {2014})}\BibitemShut {NoStop}%
\bibitem [{\citenamefont {Pirkkalainen}\ \emph {et~al.}(2013)\citenamefont
  {Pirkkalainen}, \citenamefont {Cho}, \citenamefont {Li}, \citenamefont
  {Paraoanu}, \citenamefont {Hakonen},\ and\ \citenamefont
  {Sillanp\"{a}\"{a}}}]{Pirkkalainen2013}%
  \BibitemOpen
  \bibfield  {author} {\bibinfo {author} {\bibfnamefont {J.-M.}\ \bibnamefont
  {Pirkkalainen}}, \bibinfo {author} {\bibfnamefont {S.~U.}\ \bibnamefont
  {Cho}}, \bibinfo {author} {\bibfnamefont {J.}~\bibnamefont {Li}}, \bibinfo
  {author} {\bibfnamefont {G.~S.}\ \bibnamefont {Paraoanu}}, \bibinfo {author}
  {\bibfnamefont {P.~J.}\ \bibnamefont {Hakonen}}, \ and\ \bibinfo {author}
  {\bibfnamefont {M.~A.}\ \bibnamefont {Sillanp\"{a}\"{a}}},\ }\href {\doibase
  10.1038/nature11821} {\bibfield  {journal} {\bibinfo  {journal} {Nature}\
  }\textbf {\bibinfo {volume} {494}},\ \bibinfo {pages} {211} (\bibinfo {year}
  {2013})}\BibitemShut {NoStop}%
\bibitem [{\citenamefont {Hafezi}\ \emph {et~al.}(2012)\citenamefont {Hafezi},
  \citenamefont {Kim}, \citenamefont {Rolston}, \citenamefont {Orozco},
  \citenamefont {Lev},\ and\ \citenamefont {Taylor}}]{Hafezi2012}%
  \BibitemOpen
  \bibfield  {author} {\bibinfo {author} {\bibfnamefont {M.}~\bibnamefont
  {Hafezi}}, \bibinfo {author} {\bibfnamefont {Z.}~\bibnamefont {Kim}},
  \bibinfo {author} {\bibfnamefont {S.~L.}\ \bibnamefont {Rolston}}, \bibinfo
  {author} {\bibfnamefont {L.~A.}\ \bibnamefont {Orozco}}, \bibinfo {author}
  {\bibfnamefont {B.~L.}\ \bibnamefont {Lev}}, \ and\ \bibinfo {author}
  {\bibfnamefont {J.~M.}\ \bibnamefont {Taylor}},\ }\href {\doibase
  10.1103/PhysRevA.85.020302} {\bibfield  {journal} {\bibinfo  {journal}
  {Physical Review A}\ }\textbf {\bibinfo {volume} {85}},\ \bibinfo {pages}
  {020302} (\bibinfo {year} {2012})}\BibitemShut {NoStop}%
\bibitem [{\citenamefont {Kubo}\ \emph {et~al.}(2010)\citenamefont {Kubo},
  \citenamefont {Ong}, \citenamefont {Bertet}, \citenamefont {Vion},
  \citenamefont {Jacques}, \citenamefont {Zheng}, \citenamefont {Dr\'{e}au},
  \citenamefont {Roch}, \citenamefont {Auffeves}, \citenamefont {Jelezko},
  \citenamefont {Wrachtrup}, \citenamefont {Barthe}, \citenamefont {Bergonzo},\
  and\ \citenamefont {Esteve}}]{Kubo2010}%
  \BibitemOpen
  \bibfield  {author} {\bibinfo {author} {\bibfnamefont {Y.}~\bibnamefont
  {Kubo}}, \bibinfo {author} {\bibfnamefont {F.~R.}\ \bibnamefont {Ong}},
  \bibinfo {author} {\bibfnamefont {P.}~\bibnamefont {Bertet}}, \bibinfo
  {author} {\bibfnamefont {D.}~\bibnamefont {Vion}}, \bibinfo {author}
  {\bibfnamefont {V.}~\bibnamefont {Jacques}}, \bibinfo {author} {\bibfnamefont
  {D.}~\bibnamefont {Zheng}}, \bibinfo {author} {\bibfnamefont
  {A.}~\bibnamefont {Dr\'{e}au}}, \bibinfo {author} {\bibfnamefont {J.-F.}\
  \bibnamefont {Roch}}, \bibinfo {author} {\bibfnamefont {A.}~\bibnamefont
  {Auffeves}}, \bibinfo {author} {\bibfnamefont {F.}~\bibnamefont {Jelezko}},
  \bibinfo {author} {\bibfnamefont {J.}~\bibnamefont {Wrachtrup}}, \bibinfo
  {author} {\bibfnamefont {M.~F.}\ \bibnamefont {Barthe}}, \bibinfo {author}
  {\bibfnamefont {P.}~\bibnamefont {Bergonzo}}, \ and\ \bibinfo {author}
  {\bibfnamefont {D.}~\bibnamefont {Esteve}},\ }\href {\doibase
  10.1103/PhysRevLett.105.140502} {\bibfield  {journal} {\bibinfo  {journal}
  {Physical Review Letters}\ }\textbf {\bibinfo {volume} {105}},\ \bibinfo
  {pages} {140502} (\bibinfo {year} {2010})}\BibitemShut {NoStop}%
\bibitem [{\citenamefont {Williamson}\ \emph {et~al.}(2014)\citenamefont
  {Williamson}, \citenamefont {Chen},\ and\ \citenamefont
  {Longdell}}]{Longdell2014}%
  \BibitemOpen
  \bibfield  {author} {\bibinfo {author} {\bibfnamefont {L.~A.}\ \bibnamefont
  {Williamson}}, \bibinfo {author} {\bibfnamefont {Y.-H.}\ \bibnamefont
  {Chen}}, \ and\ \bibinfo {author} {\bibfnamefont {J.~J.}\ \bibnamefont
  {Longdell}},\ }\href {\doibase 10.1103/PhysRevLett.113.203601} {\bibfield
  {journal} {\bibinfo  {journal} {Phys. Rev. Lett.}\ }\textbf {\bibinfo
  {volume} {113}},\ \bibinfo {pages} {203601} (\bibinfo {year}
  {2014})}\BibitemShut {NoStop}%
\bibitem [{\citenamefont {Tsang}(2010)}]{Tsang2010}%
  \BibitemOpen
  \bibfield  {author} {\bibinfo {author} {\bibfnamefont {M.}~\bibnamefont
  {Tsang}},\ }\href {\doibase 10.1103/PhysRevA.81.063837} {\bibfield  {journal}
  {\bibinfo  {journal} {Physical Review A}\ }\textbf {\bibinfo {volume} {81}},\
  \bibinfo {pages} {063837} (\bibinfo {year} {2010})}\BibitemShut {NoStop}%
\bibitem [{\citenamefont {Marquardt}\ \emph {et~al.}(2007)\citenamefont
  {Marquardt}, \citenamefont {Chen}, \citenamefont {Clerk},\ and\ \citenamefont
  {Girvin}}]{Marquardt2007}%
  \BibitemOpen
  \bibfield  {author} {\bibinfo {author} {\bibfnamefont {F.}~\bibnamefont
  {Marquardt}}, \bibinfo {author} {\bibfnamefont {J.}~\bibnamefont {Chen}},
  \bibinfo {author} {\bibfnamefont {A.}~\bibnamefont {Clerk}}, \ and\ \bibinfo
  {author} {\bibfnamefont {S.}~\bibnamefont {Girvin}},\ }\href {\doibase
  10.1103/PhysRevLett.99.093902} {\bibfield  {journal} {\bibinfo  {journal}
  {Physical Review Letters}\ }\textbf {\bibinfo {volume} {99}},\ \bibinfo
  {pages} {093902} (\bibinfo {year} {2007})}\BibitemShut {NoStop}%
\bibitem [{\citenamefont {Wilson-Rae}\ \emph {et~al.}(2007)\citenamefont
  {Wilson-Rae}, \citenamefont {Nooshi}, \citenamefont {Zwerger},\ and\
  \citenamefont {Kippenberg}}]{Wilson-Rae2007}%
  \BibitemOpen
  \bibfield  {author} {\bibinfo {author} {\bibfnamefont {I.}~\bibnamefont
  {Wilson-Rae}}, \bibinfo {author} {\bibfnamefont {N.}~\bibnamefont {Nooshi}},
  \bibinfo {author} {\bibfnamefont {W.}~\bibnamefont {Zwerger}}, \ and\
  \bibinfo {author} {\bibfnamefont {T.}~\bibnamefont {Kippenberg}},\ }\href
  {\doibase 10.1103/PhysRevLett.99.093901} {\bibfield  {journal} {\bibinfo
  {journal} {Physical Review Letters}\ }\textbf {\bibinfo {volume} {99}},\
  \bibinfo {pages} {093901} (\bibinfo {year} {2007})}\BibitemShut {NoStop}%
\bibitem [{\citenamefont {Schliesser}\ \emph {et~al.}(2008)\citenamefont
  {Schliesser}, \citenamefont {Rivi\`{e}re}, \citenamefont {Anetsberger},
  \citenamefont {Arcizet},\ and\ \citenamefont {Kippenberg}}]{Schliesser2008}%
  \BibitemOpen
  \bibfield  {author} {\bibinfo {author} {\bibfnamefont {A.}~\bibnamefont
  {Schliesser}}, \bibinfo {author} {\bibfnamefont {R.}~\bibnamefont
  {Rivi\`{e}re}}, \bibinfo {author} {\bibfnamefont {G.}~\bibnamefont
  {Anetsberger}}, \bibinfo {author} {\bibfnamefont {O.}~\bibnamefont
  {Arcizet}}, \ and\ \bibinfo {author} {\bibfnamefont {T.~J.}\ \bibnamefont
  {Kippenberg}},\ }\href {\doibase 10.1038/nphys939} {\bibfield  {journal}
  {\bibinfo  {journal} {Nature Physics}\ }\textbf {\bibinfo {volume} {4}},\
  \bibinfo {pages} {415} (\bibinfo {year} {2008})}\BibitemShut {NoStop}%
\bibitem [{\citenamefont {Wallraff}\ \emph {et~al.}(2004)\citenamefont
  {Wallraff}, \citenamefont {Schuster}, \citenamefont {Blais}, \citenamefont
  {Frunzio}, \citenamefont {Huang}, \citenamefont {Majer}, \citenamefont
  {Kumar}, \citenamefont {Girvin},\ and\ \citenamefont
  {Schoelkopf}}]{Wallraff2004}%
  \BibitemOpen
  \bibfield  {author} {\bibinfo {author} {\bibfnamefont {A.}~\bibnamefont
  {Wallraff}}, \bibinfo {author} {\bibfnamefont {D.~I.}\ \bibnamefont
  {Schuster}}, \bibinfo {author} {\bibfnamefont {A.}~\bibnamefont {Blais}},
  \bibinfo {author} {\bibfnamefont {L.}~\bibnamefont {Frunzio}}, \bibinfo
  {author} {\bibfnamefont {R.-S.}\ \bibnamefont {Huang}}, \bibinfo {author}
  {\bibfnamefont {J.}~\bibnamefont {Majer}}, \bibinfo {author} {\bibfnamefont
  {S.}~\bibnamefont {Kumar}}, \bibinfo {author} {\bibfnamefont {S.~M.}\
  \bibnamefont {Girvin}}, \ and\ \bibinfo {author} {\bibfnamefont {R.~J.}\
  \bibnamefont {Schoelkopf}},\ }\href {\doibase 10.1038/nature02851} {\bibfield
   {journal} {\bibinfo  {journal} {Nature}\ }\textbf {\bibinfo {volume}
  {431}},\ \bibinfo {pages} {162} (\bibinfo {year} {2004})}\BibitemShut
  {NoStop}%
\bibitem [{\citenamefont {Ilchenko}\ \emph {et~al.}(2003)\citenamefont
  {Ilchenko}, \citenamefont {Savchenkov}, \citenamefont {Matsko},\ and\
  \citenamefont {Maleki}}]{Ilchenko2003}%
  \BibitemOpen
  \bibfield  {author} {\bibinfo {author} {\bibfnamefont {V.~S.}\ \bibnamefont
  {Ilchenko}}, \bibinfo {author} {\bibfnamefont {A.~A.}\ \bibnamefont
  {Savchenkov}}, \bibinfo {author} {\bibfnamefont {A.~B.}\ \bibnamefont
  {Matsko}}, \ and\ \bibinfo {author} {\bibfnamefont {L.}~\bibnamefont
  {Maleki}},\ }\href {\doibase 10.1364/JOSAB.20.000333} {\bibfield  {journal}
  {\bibinfo  {journal} {Journal of the Optical Society of America B}\ }\textbf
  {\bibinfo {volume} {20}},\ \bibinfo {pages} {333} (\bibinfo {year}
  {2003})}\BibitemShut {NoStop}%
\bibitem [{\citenamefont {Strekalov}\ \emph {et~al.}(2009)\citenamefont
  {Strekalov}, \citenamefont {Savchenkov}, \citenamefont {Matsko},\ and\
  \citenamefont {Yu}}]{Strekalov2009}%
  \BibitemOpen
  \bibfield  {author} {\bibinfo {author} {\bibfnamefont {D.~V.}\ \bibnamefont
  {Strekalov}}, \bibinfo {author} {\bibfnamefont {A.~A.}\ \bibnamefont
  {Savchenkov}}, \bibinfo {author} {\bibfnamefont {A.~B.}\ \bibnamefont
  {Matsko}}, \ and\ \bibinfo {author} {\bibfnamefont {N.}~\bibnamefont {Yu}},\
  }\href {http://www.ncbi.nlm.nih.gov/pubmed/19282908} {\bibfield  {journal}
  {\bibinfo  {journal} {Optics letters}\ }\textbf {\bibinfo {volume} {34}},\
  \bibinfo {pages} {713} (\bibinfo {year} {2009})}\BibitemShut {NoStop}%
\bibitem [{\citenamefont {Ilchenko}\ \emph {et~al.}(2004)\citenamefont
  {Ilchenko}, \citenamefont {Savchenkov}, \citenamefont {Matsko},\ and\
  \citenamefont {Maleki}}]{Ilchenko2004}%
  \BibitemOpen
  \bibfield  {author} {\bibinfo {author} {\bibfnamefont {V.}~\bibnamefont
  {Ilchenko}}, \bibinfo {author} {\bibfnamefont {A.}~\bibnamefont
  {Savchenkov}}, \bibinfo {author} {\bibfnamefont {A.}~\bibnamefont {Matsko}},
  \ and\ \bibinfo {author} {\bibfnamefont {L.}~\bibnamefont {Maleki}},\ }\href
  {\doibase 10.1103/PhysRevLett.92.043903} {\bibfield  {journal} {\bibinfo
  {journal} {Physical Review Letters}\ }\textbf {\bibinfo {volume} {92}},\
  \bibinfo {pages} {043903} (\bibinfo {year} {2004})}\BibitemShut {NoStop}%
\bibitem [{\citenamefont {Leduc}\ \emph {et~al.}(2010)\citenamefont {Leduc},
  \citenamefont {Bumble}, \citenamefont {Day}, \citenamefont {Eom},
  \citenamefont {Gao}, \citenamefont {Golwala}, \citenamefont {Mazin},
  \citenamefont {McHugh}, \citenamefont {Merrill}, \citenamefont {Moore},
  \citenamefont {Noroozian}, \citenamefont {Turner},\ and\ \citenamefont
  {Zmuidzinas}}]{Leduc2010}%
  \BibitemOpen
  \bibfield  {author} {\bibinfo {author} {\bibfnamefont {H.~G.}\ \bibnamefont
  {Leduc}}, \bibinfo {author} {\bibfnamefont {B.}~\bibnamefont {Bumble}},
  \bibinfo {author} {\bibfnamefont {P.~K.}\ \bibnamefont {Day}}, \bibinfo
  {author} {\bibfnamefont {B.~H.}\ \bibnamefont {Eom}}, \bibinfo {author}
  {\bibfnamefont {J.}~\bibnamefont {Gao}}, \bibinfo {author} {\bibfnamefont
  {S.}~\bibnamefont {Golwala}}, \bibinfo {author} {\bibfnamefont {B.~a.}\
  \bibnamefont {Mazin}}, \bibinfo {author} {\bibfnamefont {S.}~\bibnamefont
  {McHugh}}, \bibinfo {author} {\bibfnamefont {A.}~\bibnamefont {Merrill}},
  \bibinfo {author} {\bibfnamefont {D.~C.}\ \bibnamefont {Moore}}, \bibinfo
  {author} {\bibfnamefont {O.}~\bibnamefont {Noroozian}}, \bibinfo {author}
  {\bibfnamefont {A.~D.}\ \bibnamefont {Turner}}, \ and\ \bibinfo {author}
  {\bibfnamefont {J.}~\bibnamefont {Zmuidzinas}},\ }\href {\doibase
  10.1063/1.3480420} {\bibfield  {journal} {\bibinfo  {journal} {Applied
  Physics Letters}\ }\textbf {\bibinfo {volume} {97}},\ \bibinfo {pages}
  {102509} (\bibinfo {year} {2010})}\BibitemShut {NoStop}%
\bibitem [{\citenamefont {Guarino}\ \emph {et~al.}(2007)\citenamefont
  {Guarino}, \citenamefont {Poberaj}, \citenamefont {Rezzonico}, \citenamefont
  {Degl'Innocenti},\ and\ \citenamefont {G\"{u}nter}}]{Guarino2007}%
  \BibitemOpen
  \bibfield  {author} {\bibinfo {author} {\bibfnamefont {A.}~\bibnamefont
  {Guarino}}, \bibinfo {author} {\bibfnamefont {G.}~\bibnamefont {Poberaj}},
  \bibinfo {author} {\bibfnamefont {D.}~\bibnamefont {Rezzonico}}, \bibinfo
  {author} {\bibfnamefont {R.}~\bibnamefont {Degl'Innocenti}}, \ and\ \bibinfo
  {author} {\bibfnamefont {P.}~\bibnamefont {G\"{u}nter}},\ }\href {\doibase
  10.1038/nphoton.2007.93} {\bibfield  {journal} {\bibinfo  {journal} {Nature
  Photonics}\ }\textbf {\bibinfo {volume} {1}},\ \bibinfo {pages} {407}
  (\bibinfo {year} {2007})}\BibitemShut {NoStop}%
\bibitem [{\citenamefont {Xiong}\ \emph {et~al.}(2012)\citenamefont {Xiong},
  \citenamefont {Pernice}, \citenamefont {Sun}, \citenamefont {Schuck},
  \citenamefont {Fong},\ and\ \citenamefont {Tang}}]{Xiong2012}%
  \BibitemOpen
  \bibfield  {author} {\bibinfo {author} {\bibfnamefont {C.}~\bibnamefont
  {Xiong}}, \bibinfo {author} {\bibfnamefont {W.~H.~P.}\ \bibnamefont
  {Pernice}}, \bibinfo {author} {\bibfnamefont {X.}~\bibnamefont {Sun}},
  \bibinfo {author} {\bibfnamefont {C.}~\bibnamefont {Schuck}}, \bibinfo
  {author} {\bibfnamefont {K.~Y.}\ \bibnamefont {Fong}}, \ and\ \bibinfo
  {author} {\bibfnamefont {H.~X.}\ \bibnamefont {Tang}},\ }\href {\doibase
  10.1088/1367-2630/14/9/095014} {\bibfield  {journal} {\bibinfo  {journal}
  {New Journal of Physics}\ }\textbf {\bibinfo {volume} {14}},\ \bibinfo
  {pages} {095014} (\bibinfo {year} {2012})}\BibitemShut {NoStop}%
\bibitem [{Note1()}]{Note1}%
  \BibitemOpen
  \bibinfo {note} {Which has recently become commercially available from
  NanoLN}\BibitemShut {NoStop}%
\bibitem [{\citenamefont {Wang}\ \emph {et~al.}(2015)\citenamefont {Wang},
  \citenamefont {Bo}, \citenamefont {Wan}, \citenamefont {Li}, \citenamefont
  {Gao}, \citenamefont {Li}, \citenamefont {Zhang},\ and\ \citenamefont
  {Xu}}]{Wang2015}%
  \BibitemOpen
  \bibfield  {author} {\bibinfo {author} {\bibfnamefont {J.}~\bibnamefont
  {Wang}}, \bibinfo {author} {\bibfnamefont {F.}~\bibnamefont {Bo}}, \bibinfo
  {author} {\bibfnamefont {S.}~\bibnamefont {Wan}}, \bibinfo {author}
  {\bibfnamefont {W.}~\bibnamefont {Li}}, \bibinfo {author} {\bibfnamefont
  {F.}~\bibnamefont {Gao}}, \bibinfo {author} {\bibfnamefont {J.}~\bibnamefont
  {Li}}, \bibinfo {author} {\bibfnamefont {G.}~\bibnamefont {Zhang}}, \ and\
  \bibinfo {author} {\bibfnamefont {J.}~\bibnamefont {Xu}},\ }\href {\doibase
  10.1364/OE.23.023072} {\bibfield  {journal} {\bibinfo  {journal} {Opt.
  Express}\ }\textbf {\bibinfo {volume} {23}},\ \bibinfo {pages} {23072}
  (\bibinfo {year} {2015})}\BibitemShut {NoStop}%
\bibitem [{\citenamefont {Wang}\ and\ \citenamefont {Bhave}(2014)}]{Wang2014a}%
  \BibitemOpen
  \bibfield  {author} {\bibinfo {author} {\bibfnamefont {R.}~\bibnamefont
  {Wang}}\ and\ \bibinfo {author} {\bibfnamefont {S.~A.}\ \bibnamefont
  {Bhave}},\ }\href {http://arxiv.org/abs/1409.6351} {\  (\bibinfo {year}
  {2014})},\ \Eprint {http://arxiv.org/abs/arXiv:1409.6351} {arXiv:1409.6351}
  \BibitemShut {NoStop}%
\bibitem [{\citenamefont {Poberaj}\ \emph {et~al.}(2012)\citenamefont
  {Poberaj}, \citenamefont {Hu}, \citenamefont {Sohler},\ and\ \citenamefont
  {G\"{u}nter}}]{Poberaj2012}%
  \BibitemOpen
  \bibfield  {author} {\bibinfo {author} {\bibfnamefont {G.}~\bibnamefont
  {Poberaj}}, \bibinfo {author} {\bibfnamefont {H.}~\bibnamefont {Hu}},
  \bibinfo {author} {\bibfnamefont {W.}~\bibnamefont {Sohler}}, \ and\ \bibinfo
  {author} {\bibfnamefont {P.}~\bibnamefont {G\"{u}nter}},\ }\href {\doibase
  10.1002/lpor.201100035} {\bibfield  {journal} {\bibinfo  {journal} {Laser \&
  Photonics Reviews}\ }\textbf {\bibinfo {volume} {6}},\ \bibinfo {pages} {488}
  (\bibinfo {year} {2012})}\BibitemShut {NoStop}%
\bibitem [{\citenamefont {Oxborrow}(2007)}]{Oxborrow2007}%
  \BibitemOpen
  \bibfield  {author} {\bibinfo {author} {\bibfnamefont {M.}~\bibnamefont
  {Oxborrow}},\ }in\ \href {\doibase 10.1117/12.714954} {\emph {\bibinfo
  {booktitle} {Proc. of SPIE}}},\ Vol.\ \bibinfo {volume} {6452},\ \bibinfo
  {editor} {edited by\ \bibinfo {editor} {\bibfnamefont {A.~V.}\ \bibnamefont
  {Kudryashov}}, \bibinfo {editor} {\bibfnamefont {A.~H.}\ \bibnamefont
  {Paxton}}, \ and\ \bibinfo {editor} {\bibfnamefont {V.~S.}\ \bibnamefont
  {Ilchenko}}}\ (\bibinfo {year} {2007})\ pp.\ \bibinfo {pages}
  {64520J--64520J--12}\BibitemShut {NoStop}%
\bibitem [{\citenamefont {Schwinger}(1943)}]{Schwinger1943}%
  \BibitemOpen
  \bibfield  {author} {\bibinfo {author} {\bibfnamefont {J.}~\bibnamefont
  {Schwinger}},\ }\href@noop {} {\bibfield  {journal} {\bibinfo  {journal} {MIT
  Radiation Laboratory Report no. 43-34}\ } (\bibinfo {year}
  {1943})}\BibitemShut {NoStop}%
\bibitem [{\citenamefont {Barends}\ \emph {et~al.}(2013)\citenamefont
  {Barends}, \citenamefont {Kelly}, \citenamefont {Megrant}, \citenamefont
  {Sank}, \citenamefont {Jeffrey}, \citenamefont {Chen}, \citenamefont {Yin},
  \citenamefont {Chiaro}, \citenamefont {Mutus}, \citenamefont {Neill},
  \citenamefont {O'Malley}, \citenamefont {Roushan}, \citenamefont {Wenner},
  \citenamefont {White}, \citenamefont {Cleland},\ and\ \citenamefont
  {Martinis}}]{Barends2013}%
  \BibitemOpen
  \bibfield  {author} {\bibinfo {author} {\bibfnamefont {R.}~\bibnamefont
  {Barends}}, \bibinfo {author} {\bibfnamefont {J.}~\bibnamefont {Kelly}},
  \bibinfo {author} {\bibfnamefont {A.}~\bibnamefont {Megrant}}, \bibinfo
  {author} {\bibfnamefont {D.}~\bibnamefont {Sank}}, \bibinfo {author}
  {\bibfnamefont {E.}~\bibnamefont {Jeffrey}}, \bibinfo {author} {\bibfnamefont
  {Y.}~\bibnamefont {Chen}}, \bibinfo {author} {\bibfnamefont {Y.}~\bibnamefont
  {Yin}}, \bibinfo {author} {\bibfnamefont {B.}~\bibnamefont {Chiaro}},
  \bibinfo {author} {\bibfnamefont {J.}~\bibnamefont {Mutus}}, \bibinfo
  {author} {\bibfnamefont {C.}~\bibnamefont {Neill}}, \bibinfo {author}
  {\bibfnamefont {P.}~\bibnamefont {O'Malley}}, \bibinfo {author}
  {\bibfnamefont {P.}~\bibnamefont {Roushan}}, \bibinfo {author} {\bibfnamefont
  {J.}~\bibnamefont {Wenner}}, \bibinfo {author} {\bibfnamefont {T.~C.}\
  \bibnamefont {White}}, \bibinfo {author} {\bibfnamefont {A.~N.}\ \bibnamefont
  {Cleland}}, \ and\ \bibinfo {author} {\bibfnamefont {J.~M.}\ \bibnamefont
  {Martinis}},\ }\href {\doibase 10.1103/PhysRevLett.111.080502} {\bibfield
  {journal} {\bibinfo  {journal} {Phys. Rev. Lett.}\ }\textbf {\bibinfo
  {volume} {111}},\ \bibinfo {pages} {080502} (\bibinfo {year}
  {2013})}\BibitemShut {NoStop}%
\bibitem [{\citenamefont {Tsang}(2011)}]{Tsang2011}%
  \BibitemOpen
  \bibfield  {author} {\bibinfo {author} {\bibfnamefont {M.}~\bibnamefont
  {Tsang}},\ }\href {\doibase 10.1103/PhysRevA.84.043845} {\bibfield  {journal}
  {\bibinfo  {journal} {Physical Review A}\ }\textbf {\bibinfo {volume} {84}},\
  \bibinfo {pages} {043845} (\bibinfo {year} {2011})}\BibitemShut {NoStop}%
\bibitem [{\citenamefont {Dobrindt}\ and\ \citenamefont
  {Kippenberg}(2010)}]{Dobrindt2010}%
  \BibitemOpen
  \bibfield  {author} {\bibinfo {author} {\bibfnamefont {J.~M.}\ \bibnamefont
  {Dobrindt}}\ and\ \bibinfo {author} {\bibfnamefont {T.~J.}\ \bibnamefont
  {Kippenberg}},\ }\href {\doibase 10.1103/PhysRevLett.104.033901} {\bibfield
  {journal} {\bibinfo  {journal} {Physical Review Letters}\ }\textbf {\bibinfo
  {volume} {104}},\ \bibinfo {pages} {033901} (\bibinfo {year}
  {2010})}\BibitemShut {NoStop}%
\bibitem [{\citenamefont {Barzanjeh}\ \emph {et~al.}(2015)\citenamefont
  {Barzanjeh}, \citenamefont {Guha}, \citenamefont {Weedbrook}, \citenamefont
  {Vitali}, \citenamefont {Shapiro},\ and\ \citenamefont
  {Pirandola}}]{Pirandola2015}%
  \BibitemOpen
  \bibfield  {author} {\bibinfo {author} {\bibfnamefont {S.}~\bibnamefont
  {Barzanjeh}}, \bibinfo {author} {\bibfnamefont {S.}~\bibnamefont {Guha}},
  \bibinfo {author} {\bibfnamefont {C.}~\bibnamefont {Weedbrook}}, \bibinfo
  {author} {\bibfnamefont {D.}~\bibnamefont {Vitali}}, \bibinfo {author}
  {\bibfnamefont {J.~H.}\ \bibnamefont {Shapiro}}, \ and\ \bibinfo {author}
  {\bibfnamefont {S.}~\bibnamefont {Pirandola}},\ }\href {\doibase
  10.1103/PhysRevLett.114.080503} {\bibfield  {journal} {\bibinfo  {journal}
  {Phys. Rev. Lett.}\ }\textbf {\bibinfo {volume} {114}},\ \bibinfo {pages}
  {080503} (\bibinfo {year} {2015})}\BibitemShut {NoStop}%
\end{thebibliography}%

\end{document}